\documentclass[pra,showpacs,twocolumn,epsfig]{revtex4}
\usepackage{amsfonts}
\usepackage{amsmath}
\usepackage{amssymb}
\usepackage{graphicx}

\begin{document}

\title{Spin squeezing and maximal-squeezing time}
\author{Guang-Ri Jin \cite{email} and Sang Wook Kim \cite{email1}}
\affiliation{Department of Physics Education and Department of
Physics, Pusan National University, Busan 609-735, Korea}
\date{\today }

\begin{abstract}
Spin squeezing of a nonlinear interaction model with
Josephson-like coupling is studied to obtain time scale of maximal
squeezing. Based upon two exactly solvable cases for two and three
particles, we find that the maximal-squeezing time depends on the
level spacing between the ground state and its next neighbor
eigenstate.
\end{abstract}

\maketitle


\section{INTRODUCTION}

The phenomenon of spin squeezing in collective spin system have
attracted much attention for decades not only because of fundamental
physical interests
\cite{Kitagawa,Wineland,Wineland1,Wineland2,Wineland3,Kuzmich,Hald,Geremia,Rojo},
but also for its possible application in atomic clocks for reducing
quantum noise \cite{Wineland,Wineland1,Wineland2,Wineland3} and
quantum information \cite{Sorensen,You,Wang,Lewenstein,Yi}. The
occurrence of spin squeezing is due to quantum correlations among
individual spins, which requires at least two spins and nonlinear
interaction between them. Kitagawa and Uea have studied the spin
squeezing generated by the so-called one-axis twisting (OAT) model
with Hamiltonian: $\hat{H}_{\text{OAT}}=2\kappa \hat{J}_{z}^{2}$ \cite%
{Kitagawa}. Possible realization of the OAT-type squeezing in a
two-component Bose-Einstein Condensate (TBEC)
\cite{Sorensen,Molmer}, and atomic ensemble system in a dispersive
regime \cite{Takeuchi} have been investigated recently. S\o rensen
et al. also proposed that the spin squeezing can be used as a
measure of many-particle quantum entanglement \cite{Sorensen}.

So far the OAT-type spin squeezing was mainly studied in
Heisenberg picture. As a result, the explicit expression of the
spin squeezed state is unknown. Moreover, the direction that spin
squeezing is observed varies with time \cite{You}. Jaksch et al.
have shown that the OAT-type SSS can be
stored for arbitrarily long time by removing the self-interaction \cite%
{Jaksch}. However, it might not be easy to handle in experiment since the
precisely designed additional pulses are crucially required. In Refs. \cite%
{Law,Bigelow}, the authors proposed the constant-coupling scheme by
introducing additional Josephson-like coupling $\Omega \hat{J}_{x}$
to the OAT model. It was shown that the Josephson interaction
results in an enhancement of spin squeezing compared with that of
the OAT. Moreover, the strongest squeezing appears in the $z$
direction \cite{Law}, which, however, is valid only at the
maximal-squeezing time (MST). Although some formulas of the MST for
extremely small \cite{Kitagawa} or large coupling
\cite{Law,Bigelow,Jenkins,Choi,Liang} have already been known, it is
challenging to determine the MST within an intermediate coupling
$1<\Omega/\kappa<<N$ [where $N$ is total particle number].

In this paper, we reconsider the constant-coupling scheme
\cite{Law,Bigelow} with the purpose to determine the MST. We find
all the analytic solutions for two- and three-particle cases.
Motivated by the exactly solvable cases, we show that the MST
depends on the level spacing between the ground state and its next
neighbor eigenstate. We explain it by investigating the spectral
distribution of the spin state, and find only the two lowest
available levels are predominantly occupied. Our paper is
organized as follows. In Sec. II, we introduce theoretical model
and derive some basic formulas. To proceed, in Sec. III, we gives
some analytic expressions for the cases of $N=2$ and $N=3$. In
Sec. IV, we study the spin squeezing for many-particle cases, and
present exact diagonalization method to obtain the MST. Moreover,
we compare our result with its analytic solution. Finally, a
summary of our paper is presented.

\section{Theoretical model}

Formally, a two-level atom can be regarded as a fictitious
spin-1/2 particle with spin operators $s_{z}^{(i)}=(|b
\rangle_{ii}\langle b|-|a\rangle_{ii}\langle a|)/2$, and
$s_{+}^{(i)}=(s_{-}^{(i)})^{\dagger}=|b\rangle_{ii}\langle a|$,
where $|a\rangle_{i}$ and $|b\rangle_{i}$ are the internal states
of the $i$th atom. We consider an ensemble of $N$ atoms with its
dynamics described by collective spin operator:
$\hat{J}=\sum_{i=1}^{N} s^{(i)}$. The spin squeezing is quantified
by a parameter \cite{Kitagawa}:
\begin{equation}
\xi =\frac{\sqrt{2}(\Delta \hat{J}_{\mathbf{n}})_{\min}}{j^{1/2}},
\label{xi}
\end{equation}%
where $j=N/2$, and $(\Delta \hat{J}_{\mathbf{n}})_{\min}$ represents
the smallest variance of a spin component
$\hat{J}_{\mathbf{n}}=\hat{J}\cdot \mathbf{n}$ normal to the mean
spin $\langle \hat{J}\rangle $. For a coherent spin state (CSS), the
variance $(\Delta \hat{J}_{\mathbf{n}})_{\min}=\sqrt{j/2}$ and
$\xi=1$. In general, a spin state is called spin squeezed state
(SSS) if the variance of the spin component $\hat{J}_{\mathbf{n}}$
is smaller than that of the CSS, i.e. $\xi <1$.

Follow Refs.
\cite{Milburn,Smerzi,Villain,TMA1,TMA2,TMA3,Savage99,Kuang}, we
consider a nonlinear spin system governed by
\begin{equation}
\hat{H}=\Omega \hat{J}_{x}+2\kappa \hat{J}_{z}^{2},  \label{hamiltonian}
\end{equation}%
which can be realized in the TBEC \cite{Hall,Stenger}. The first
term is Josephson-like coupling induced by a microwave (radio
frequency) field. The Rabi frequency $\Omega$ can be controlled by
the strength of the external field. The second term is the
self-interaction aroused from
nonlinear collision between atoms. An initial coherent spin state $%
|j,-j\rangle_{x}=e^{-i\pi J_{y}/2}|j,-j\rangle$ will be considered
in this paper. Physically, the Dicke state $|j,-j\rangle$ represents
all the atoms occupying in the internal ground state $|a\rangle$. By
applying a short $\pi /2$ pulse to the Dicke state, one can obtain
the CSS with each spin to be aligned along the negative $x$
direction \cite{Sorensen}. After that, one switches on the
Josephson-like immediately, then dynamics of the spin system is
governed by the Hamiltonian (\ref{hamiltonian}). Note that, we will
consider only positive $\kappa$ case. However, our results keep
valid in the
opposite case by using initial maximum weight state of $\hat{J}_x$, i.e., $%
|j,j\rangle_x$.

The state vector at any time $t$ can be expanded in terms of eigenstates of $%
\hat{J}_{z}$: $|\psi (t)\rangle =\sum_{m}c_{m}(t)\left\vert j,m\right\rangle
$, where $-j\leq m\leq j$. The probability amplitudes $c_{m}(t)$ can be
solved by time-dependent Schr\"{o}dinger equation, obeying
\begin{equation}
i\dot{c}_{m}=E_{m}c_{m}+X_{m}c_{m-1}+X_{-m}c_{m+1},  \label{eq_motion}
\end{equation}%
where $E_{m}=2\kappa m^{2}$, and
$X_{m}=\frac{\Omega}{2}\sqrt{(j+m)(j-m+1)}$ with $X_{-j}=0$ and
$X_{\pm m}=X_{\mp m+1}$. The probability amplitudes of
the initial CSS
\begin{equation}
c_{m}(0)=\frac{(-1)^{j+m}}{2^{j}}\sqrt{\frac{(2j)!}{%
(j-m)!(j+m)!}},\label{the initial condition}
\end{equation}
satisfy $c_{-m}(0)=c_{m}(0)$ for even $N$, and $%
c_{-m}(0)=-c_{m}(0)$ for odd $N$. Due to the symmetry properties
of the elements $X_{\pm m}$ and the initial amplitudes $c_{m}(0)$,
we obtain simple
expressions: $c_{-m}(t)=\pm c_{m}(t)$, which in turn result in $\langle \hat{%
J}_{y}\rangle =\langle \hat{J}_{z}\rangle =0$, and $\langle \hat{J}%
_{x}\rangle \neq 0$, i.e., the mean spin $\langle \hat{J}\rangle $ is always
along the $x$ axis. The spin component normal to the mean spin is $\hat{J}_{%
\mathbf{n}}=\hat{J}_{y}\sin \theta +\hat{J}_{z}\cos \theta $ and its
variance is $(\Delta \hat{J}_{\mathbf{n}})^{2}=\langle \hat{J}_{\mathbf{n}%
}^{2}\rangle -\langle \hat{J}_{\mathbf{n}}\rangle ^{2}\equiv \frac{1}{2}C+%
\frac{\cos 2\theta }{2}A+\frac{\sin 2\theta }{2}B$, where $A=\langle \hat{J}%
_{z}^{2}-\hat{J}_{y}^{2}\rangle $, $B=\langle \hat{J}_{z}\hat{J}_{y}+\hat{J}%
_{y}\hat{J}_{z}\rangle $, and $C=\langle \hat{J}_{z}^{2}+\hat{J}%
_{y}^{2}\rangle $. By minimizing the variance $(\Delta \hat{J}_{\mathbf{n}%
})^{2}$ with respect to $\theta $, we get the squeezing angle:
\begin{equation}
\theta _{\min }=\frac{1}{2}\tan ^{-1}(B/A),  \label{theta}
\end{equation}%
and the smallest variance
\begin{equation}
(\Delta \hat{J}_{\mathbf{n}})_{\min }^{2}=\frac{1}{2}C-\frac{1}{2}\sqrt{%
A^{2}+B^{2}},  \label{variance}
\end{equation}%
from which one also obtain the squeezing parameter Eq. (\ref{xi}).
We consider the spin squeezing in the intermediate coupling regime,
namely $1<\Omega /\kappa <<N$, where no analytic solutions are
available for the nonlinear spin system \cite{Law,Agarwal}. However,
we can exactly solve two- and three-particle cases. Some of
important physics can be extended to many-particle cases.

\section{Exact solvable cases}

In this section, we study the spin squeezing based on two exact
solvable cases with $N=2$ and $N=3$. Though simple, it is of general
interest to investigate the relationship between spin squeezing and
quantum entanglement
\cite{Sorensen,You,Wang,Lewenstein,ZhouL,Hagley,Messikh,ZengB}. Such
a relationship for two-particle (two-qubit)
\cite{Wineland1,Wineland2,Hagley,Messikh} and three-particle
\cite{Wineland3,ZengB} have been studied recently. Here, we focus on
dynamical behavior of the spin system to show the conditions of the
optimal squeezing and its time scale.

\subsection{Two-particle case}

\begin{figure}[tph]
\begin{centering}
\includegraphics[width=8cm]{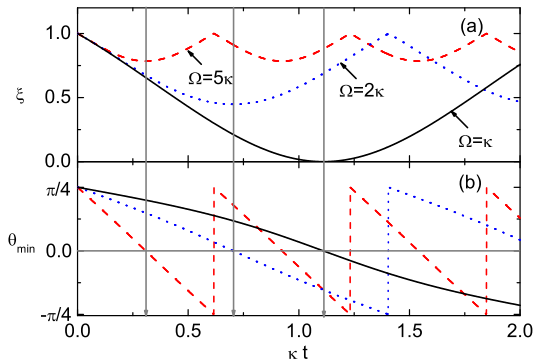} 
\end{centering}
\caption{(color online) Time evolution of (a) the squeezing
parameter and
(b) the squeezing angle for $N=2$ and various Rabi frequencies: $\Omega =5%
\protect\kappa$ (Dashed red lines), $\Omega =2\protect\kappa$ (
Dotted blue lines), and $\Omega =\protect\kappa $ (Solid black
lines). The maximal-squeezing times $t_{0}$ for different $\Omega$
are indicated by the vertical lines.} \label{fig1}
\end{figure}

For the simplest case $N=2$ ($j=1$), only three spin projections ($m=-1,$ $0,
$ $+1)$ are involved. From Eq. (\ref{eq_motion}), we obtain%
\begin{equation}
i\left(
\begin{array}{c}
\dot{p}_{0}^{(+)} \\
\dot{p}_{1}^{(+)}%
\end{array}%
\right) =\left(
\begin{array}{cc}
E_{0} & 2X_{1} \\
X_{1} & E_{1}%
\end{array}%
\right) \left(
\begin{array}{c}
p_{0}^{(+)} \\
p_{0}^{(+)}%
\end{array}%
\right),   \label{p2}
\end{equation}%
where $E_m$ and $X_m$ are defined in Eq.~(\ref{eq_motion}), and we
have introduced the linear combinations of the probability
amplitudes $p_{1}^{(+)}(t)=c_{1}(t)+c_{-1}(t)$ and $p_{0}^{(+)}(t)=2c_{0}(t)$%
, with the initial conditions $p_{1}^{(+)}(0)=1$ and $p_{0}^{(+)}(0)=-\sqrt{2%
}$. Similarly, we also introduce $p_{1}^{(-)}(t)=c_{1}(t)-c_{-1}(t)$.
However, its solution $p_{1}^{(-)}(t)=e^{-i2\kappa t}p_{1}^{(-)}(0)\equiv 0$
due to $p_{1}^{(-)}(0)=0$. Therefore, we obtain $c_{1}(t)=c_{-1}(t)\equiv
p_{1}^{(+)}/2$, which gives $\langle \hat{J}_{z}\rangle
=|c_{+1}|^{2}-|c_{-1}|^{2}\equiv 0$ and $\langle \hat{J}_{+}\rangle =\sqrt{2}%
(c_{-1}c_{0}^{\ast }+c_{0}c_{1}^{\ast })=2\sqrt{2}{\text{Re}}%
(c_{0}c_{1}^{\ast })$. Since $\langle \hat{J}_{+}\rangle $ is a real
function, $\langle \hat{J}_{y}\rangle =0$ and $\langle \hat{J}_{x}\rangle
\neq 0$, which show that the mean spin is always along the $x$ direction.
Such a result is valid for arbitrary even $N$. Eq. (\ref{p2}) can be solved
exactly, then one obtain immediately the reduced variance
\begin{equation}
(\Delta \hat{J}_{\mathbf{n}})_{\min }^{2}=\frac{1}{2}-\frac{\kappa }{S}%
\left\vert \sin St\right\vert \sqrt{1-\frac{\kappa ^{2}}{S^{2}}\sin ^{2}St},
\label{xi2}
\end{equation}%
and the squeezing angle
\begin{equation}
\theta _{\min }=\frac{1}{2}\tan ^{-1}\left[ \frac{S\cos (St)}{\Omega \sin
(St)}\right] ,  \label{theta2}
\end{equation}%
where $2S=\mathcal{E}_{3}-\mathcal{E}_{1}=2\sqrt{\Omega ^{2}+\kappa
^{2}}$ is the level spacing between the second excited state
$\mathcal{E}_{3}$ and the ground state $\mathcal{E}_{1}$, obtained
by solving the eigenvalues of the coefficient matrix of Eq.
(\ref{p2}). As shown in Fig.~\ref{fig1}(a), we find that at the
times $t_{k}^{\ast }=k\pi/S$, $\xi$ revives periodically to its
initial value $1$. In fact, apart from a globe phase, the states at
$t_{k}^{\ast}$, $\left\vert \psi (t_{k}^{\ast })\right\rangle
=(-1)^{k}e^{-i\kappa t_{k}^{\ast }}\left\vert 1,-1\right\rangle
_{x}$, are just the initial CSS.

From Eq.~(\ref{theta2}), we find that the vanishing $\theta _{\min}$
occurs at $t_{k}=(k+1/2)\pi/S$, and the state vector at $t_k$ reads
\begin{eqnarray}
\left\vert \psi (t_{k})\right\rangle &=&(-1)^{k}ie^{-i\kappa t_{k}}  \notag
\\
&&\times \left\{ \sin (\eta )\left\vert 1,-1\right\rangle _{x}-\cos (\eta
)\left\vert 1,+1\right\rangle _{x}\right\} ,  \label{EntanleS2}
\end{eqnarray}
which correspond to a superposition of two coherent spin states,
$\vert 1,-1\rangle _{x}$ and $\vert 1,+1\rangle _{x}$ with the
mixing angle $\eta =\tan ^{-1}(\Omega /\kappa )$. Obviously, if
the coupling is very strong ($\Omega \gg \kappa$), $\sin (\eta
)=\Omega /S\rightarrow 1$ and $\cos (\eta )=\kappa /S\rightarrow
0$, so $\vert \psi (t_{k})\rangle\rightarrow\vert 1,-1\rangle
_{x}$, which in turn leads to a very weak squeezing at $t_k$. On
the other hand, if the coupling is very weak ($\Omega \ll
\kappa$), $\left\vert \psi
(t_{k})\right\rangle\rightarrow\vert1,1\rangle _{x}$, which also
results in a weak squeezing at $t_k$. Therefore, we will study the
spin squeezing within the intermediate coupling regime.

In Fig.~\ref{fig1}, time evolutions of $\xi$ and $\theta_{\min}$ are
investigated for the coupling $\Omega\geq\kappa$. We observe that
local minima of $\xi$ together with $\theta_{\min}=0$ also occurs
periodically at the times $t_k$. Moreover, with the decrease of
$\Omega$, the squeezing parameter at $t_{k}$ becomes small, i.e.,
more squeezed. For the coupling $\Omega =\kappa$, the spin system is
optimally squeezed at $t_k$, as shown by the solid black lines of
Fig.~\ref{fig1}. In this case $\sin (\eta)=\cos (\eta)=1/\sqrt{2}$,
and the spin states at $t_k$ are
\begin{eqnarray}
|\psi (t_{k})\rangle &=&(-1)^{k}\frac{ie^{-i\kappa t_{k}}}{\sqrt{2}}\left\{
\left\vert 1,-1\right\rangle _{x}-\left\vert 1,+1\right\rangle _{x}\right\}
\notag \\
&=&i(-1)^{k+1}e^{-i\kappa t_{k}}\vert j=1,m=0\rangle.
\label{MSS2}
\end{eqnarray}%
Here, the state $(\vert 1,-1\rangle _{x}-\vert 1,+1\rangle _{x})/%
\sqrt{2}$ is maximally entangled (or Bell) state, while the Dicke
state $\vert j=1,m=0\rangle$ is maximally squeezed state
\cite{Wineland}. For this state, both the mean spin
$\langle\hat{J}_x\rangle$ and the variance $(\Delta
\hat{J}_{\mathbf{n}})_{\min}$ are equal to zero, which makes it hard
to define $\xi$ as Eq.~(\ref{xi}). To avoid this problem, Wineland
et al. proposed another definition of the squeezing parameter,
namely $\xi\rightarrow(j/|\langle\hat{J}\rangle|)\xi$, which gives
the smallest squeezing $1/\sqrt{2}$ for $N=2$ case \cite{Wineland}.

\subsection{Three-particle case}

For $N=3$ ($j=3/2$) case, we introduce the linear combinations of
the amplitudes $p_{m}^{(+)}(t)=c_{m}(t)+c_{-m}(t)$ with $m=1/2,3/2$.
Since $p_{3/2}^{(+)}(0)=p_{1/2}^{(+)}(0)=0$, we get
$p_{m}^{(+)}(t)=0$. Therefore, the amplitudes obey
$c_{m}(t)=-c_{-m}(t)$, from which we can prove that the mean spin is
always along the $x$ direction. Such result keeps valid for any odd
$N$ case. From Eq. (\ref{eq_motion}), we obtain a coupled equations
for the linear combinations $p_{m}^{(-)}(t)=c_{m}(t)-c_{-m}(t)$:
\begin{equation}
i\left(
\begin{array}{c}
\dot{p}_{1/2}^{(-)} \\
\dot{p}_{3/2}^{(-)}%
\end{array}%
\right) =\left(
\begin{array}{cc}
E'_{1/2} & X_{3/2} \\
X_{3/2} & E_{3/2}%
\end{array}%
\right) \left(
\begin{array}{c}
p_{1/2}^{(-)} \\
p_{3/2}^{(-)}%
\end{array}%
\right),   \label{p3}
\end{equation}%
where $E'_{1/2}=E_{1/2}-X_{1/2}$. The initial conditions are $p_{3/2}^{(-)}(0)=-1/\sqrt{2}$ and $%
p_{1/2}^{(-)}(0)=\sqrt{3/2}$. Dynamical evolution of the three-spin
system is determined solely by Eq. (\ref{p3}). The analytic
expression of the variance is
\begin{eqnarray}
(\Delta \hat{J}_{\mathbf{n}})_{\min }^{2} &=&\frac{3}{4}+\frac{3\kappa ^{2}}{%
S^{2}}\sin ^{2}St  \notag \\
&&-\frac{3\kappa }{S}\left\vert \sin St\right\vert \sqrt{1-\frac{3\kappa ^{2}%
}{S^{2}}\sin ^{2}St},  \label{xi3}
\end{eqnarray}%
and the squeezing angle is
\begin{equation}
\theta _{\min }=\frac{1}{2}\tan ^{-1}\left[ \frac{S\cos (St)}{(\Omega
+\kappa )\sin (St)}\right] ,  \label{theta3}
\end{equation}%
where $2S=\mathcal{E}_{3}-\mathcal{E}_{1}=2\sqrt{\Omega
^{2}+2\kappa \Omega +4\kappa ^{2}}$ is the level spacing for $N=3$ case, and $\mathcal{E}_{3}$ and $\mathcal{E}%
_{1}$ correspond to two eigenvalues of the coefficient matrix of
Eq.~(\ref{p3}).

\begin{figure}[tph]
\begin{centering}
\includegraphics[width=8cm]{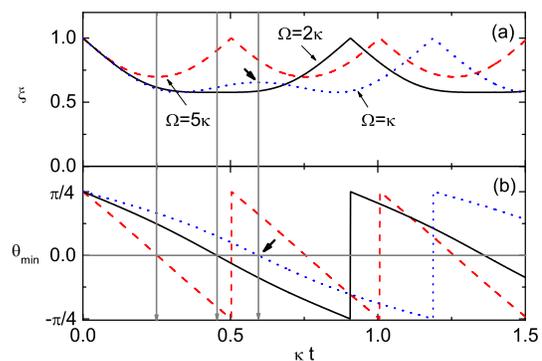} 
\end{centering}
\caption{(Color online) Time evolution of (a) the squeezing
parameter, and (b) the squeezing angle for $N=3$ case with various
Rabi frequencies: $\Omega =5\protect\kappa$ (Dashed red lines),
$\Omega =2\kappa$ (Solid black lines), and $\Omega =\protect\kappa$
(Dotted blue lines).} \label{fig2}
\end{figure}

In Fig.~\ref{fig2}, we investigate time evolution of $\xi$ and
$\theta_{\min}$ for $N=3$ case. Similar with previous $N=2$ case,
our results show that for $\Omega \geq 2\kappa$, local minima of
$\xi$ together with $\theta _{\min}=0$ also occur at the times
$St_{k}=(k+1/2)\pi$. As shown by the solid black line of
Fig.~\ref{fig2}, we find that the optimal squeezing can be obtained
at $t_k$ for the coupling $\Omega =2\kappa$. The maximally squeezed
state at $t_k$ reads
\begin{equation}
|\psi (t_{k})\rangle =\frac{i(-1)^k}{\sqrt{2}}e^{-3i\kappa t_{k}/2}\left(
\left\vert \frac{3}{2},\frac{1}{2}\right\rangle -\left\vert \frac{3}{2},-%
\frac{1}{2}\right\rangle \right).  \label{MSS3}
\end{equation}%
Such a state gives the smallest squeezing parameter
$\xi(t_k)=1/\sqrt{3}$ that the three-particle system can reach. It
is worth mentioning that for $\Omega <2\kappa$, the vanishing
$\theta_{\min}$ appearing at $t_{k}$ no longer corresponds to local
minima of $\xi$, as shown by the dotted blue lines of
Fig.~\ref{fig2}. The time scale $t_k$ is relevant to determine the
MST only for $\Omega$ equal or larger than the optimal coupling.

In short, we find some basic features for two exactly solvable
cases. Local minima of $\xi$ with $\theta_{\min}=0$ occur at the
MST $t_{k}$. This is no longer true if $\Omega$ smaller than the
optimal coupling. The time scale $t_{k}$ depends on the level
spacing $2S$ between the ground state and the second excited
state. Due to the symmetric properties of the spin system, the
first excited eigenstate is an \textit{idle level} (see below).
This is also the reason why we can introduce the linear
combinations of the amplitudes $p_m^{(\pm)}$, with $m=0, 1$ for
$N=2$, and $m=1/2, 3/2$ for $N=3$. For the optimal coupling, the
spin system will be evolved into the maximally squeezed state at
$t_k$ : $|1,0\rangle$ (for $N=2$) or $(|3/2, 1/2\rangle-|3/2,
-1/2\rangle)/\sqrt{2}$ (for $N=3$), which is just the ground state
of $2\kappa\hat{J}_z^2$. We will extend the above results to
many-particle cases.

\section{Many-particle cases: The maximal-squeezing time}

\begin{figure}[tbph]
\begin{center}
\includegraphics[width=8cm]{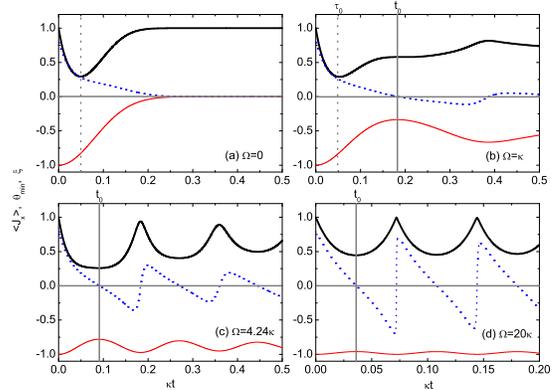}
\end{center}
\caption{(color online) Time evolution of $\protect\xi$ (thick
solid lines), $\protect\theta_{\min}$ (dashed blue lines) and
$\langle\hat{J}_x\rangle/j$
(thin red lines) for $N=40$ ($j=20$) and various Rabi frequencies: (a) $%
\Omega =0$, (b) $\Omega =\protect\kappa$, (c) $\Omega
=4.24\protect\kappa$
(the optimal coupling), and (d) $\Omega =20\protect\kappa$. The time scale $%
t_{0}$ for different $\Omega$ are indicated by the vertical lines.
} \label{fig3}
\end{figure}

In this section, we study the spin squeezing for many-particle cases
focusing on the time scale of the maximal squeezing. For
instance, we consider the spin system with particle number $N=40$ \cite%
{Kitagawa,Takeuchi}. The numerical results are shown in
Fig.~\ref{fig3}. We find that with the increase of $%
\Omega$, the squeezing $\xi$ and the mean spin
$\langle\hat{J}_x\rangle$ show collapsed oscillations
\cite{Agarwal,Jin04}. Local maxima of the mean spin
$\langle\hat{J}_x\rangle$ always appear together with the vanishing
$\theta_{\min}$. We can prove this from Heisenberg equation of
$\hat{J}_x$ and Eq.~(\ref{theta}): $d\langle\hat{J}_x\rangle/dt\sim
\langle\hat{J}_{z} \hat{J}_{y}+\hat{J}_{y}\hat{J}_{z}\rangle\sim
A\tan(2\theta_{\min})$. If the mean spin reaches its local maximum
at a certain time $t_0$, then
$\left.d\langle\hat{J}_x\rangle/dt\right|_{t_0}=0$, which leads to
$\theta_{\min}=0$ at $t_0$ provided that $A\neq0$.

As shown in Fig.~\ref{fig3}(b), for a small coupling with
$\Omega=\kappa$, there are two time scales: $t_0$ for the vanishing
$\theta_{\min}$, and $\tau_0$ for the maximal squeezing. Note that
the latter time scale $\tau_0$ closes to that of the OAT result
($\Omega=0$ case), i.e. $\kappa\tau_0\simeq0.04986$. With the
increase of $\Omega$, these two time scales become coincident, as
shown in Fig.~\ref{fig3}(c) and (d). Unlike to the exactly solvable
cases, we find that the optimal coupling for $N=40$ is not a fixed
value but can be arbitrary $\Omega$ in a region
$4.239\leq\Omega_R/\kappa\leq4.242$. Fig.~\ref{fig3}(c) represents
the optimal squeezing case with the coupling $\Omega=4.24\kappa$.
Starting from the initial CSS, the spin system evolves into the
maximally squeezed state at $\kappa t_0=0.09$.

\begin{figure}[htbp]
\begin{center}
\includegraphics[width=6cm, height=8.5cm, angle=270]{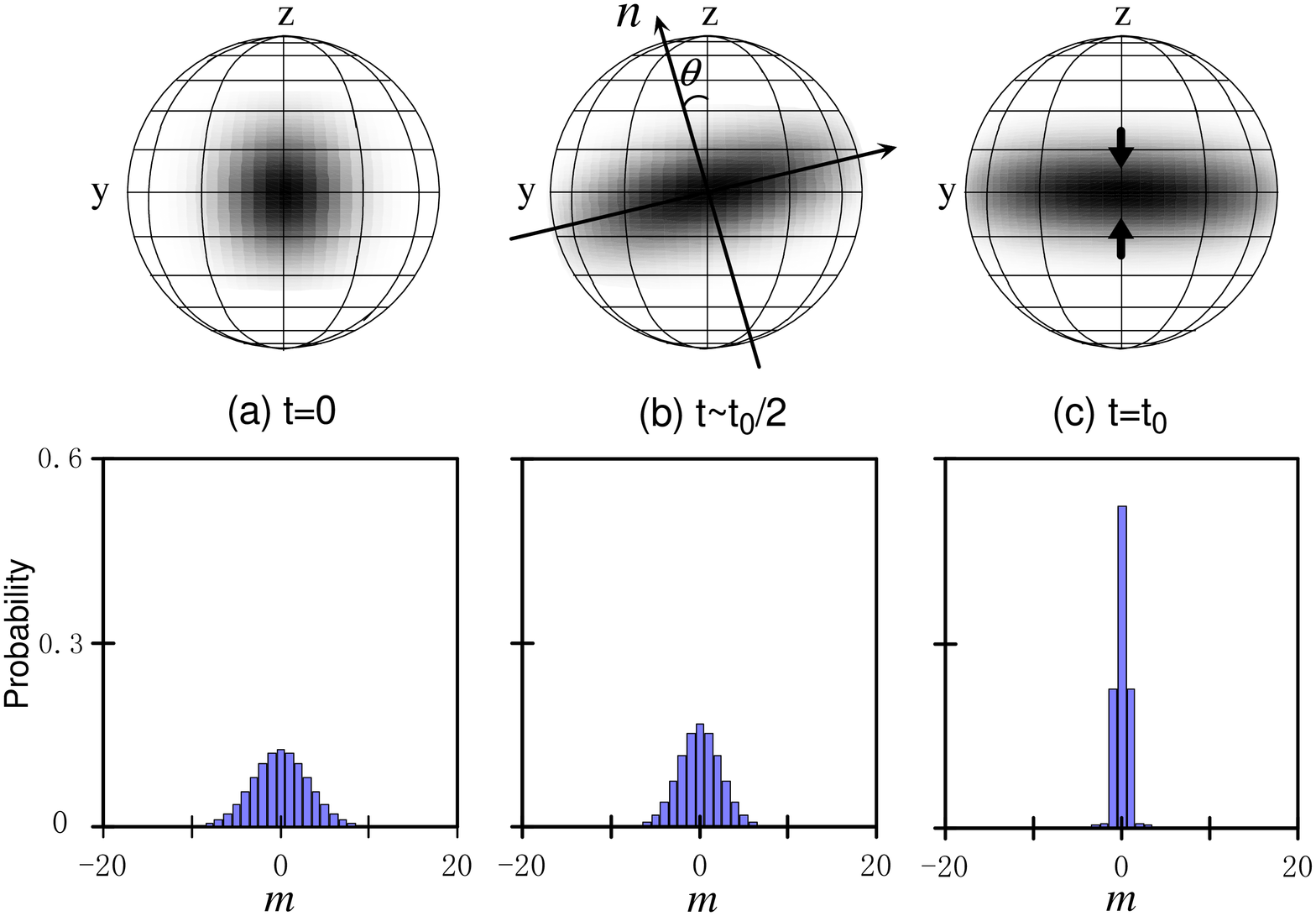}
\end{center}
\caption{Time evolution of the quasiprobability distribution $Q(\protect%
\theta, \protect\phi)$ (top) and the probability distribution
$|c_m|^2$ (bottom) at the times (a) $\protect\kappa t=0$, (b)
$\protect\kappa t=0.04$, and (c) $\protect\kappa t=0.09$ (the
MST). The QPD is normalized such that
$Q(\protect\pi/2,\protect\pi)=1$. In
(b), the spin component normal to the $x$ axis is defined as $\hat{J}_{%
\mathbf{n}}=\hat{J}\cdot\mathbf{n}$ with $\mathbf{n}=(0, \sin\protect\theta,
\cos\protect\theta) $. Other parameters are taken as those of Fig.~\protect\ref%
{fig3}(c).}
\label{fig4}
\end{figure}

To investigate the maximally SSS at $t_0$, we calculate the
quasiprobability distribution (QPD, or the Husimi function)
$Q(\theta, \phi)$ on the Bloch sphere \cite{Kitagawa}
\begin{equation}
Q(\theta, \phi)=|\langle \theta, \phi|\psi(t)\rangle|^2,
\label{quasiprobability}
\end{equation}
where $|\theta, \phi\rangle=\exp\{-i\theta(\hat{J}_x\sin\phi-\hat{J}%
_y\cos\phi)\}|j,-j\rangle$ is the generalized coherent spin state \cite%
{CSS1,CSS2}. The initial state is a particular case of the CSS, namely $%
|j,-j\rangle_x=|\theta=\pi/2, \phi=\pi\rangle$. The QPD can be
used to simulate the variation of spin uncertainties. The circle in Fig.~\ref%
{fig4}(a) represents an isotropic spin variance for the initial
CSS, while the shaded ellipse parts in Fig.~\ref%
{fig4}(b) and (c) are that of the SSS at times about $t_0/2$ and
$t_0$, respectively. Unlike to the OAT result
\cite{Kitagawa,Takeuchi}, the maximal variance reduction appears
along the $z$ axis with $\theta_{\min}=0$ \cite{Law}. In
Fig.~\ref{fig4}, we also calculate the probability distribution
$|c_m|^2=|\langle j,m|\psi(t)\rangle|^2$ of the spin state for
$N=40$ and the optimal coupling $\Omega=4.24\kappa$. Compared with
the initial CSS, we find that the maximally SSS at $t_0$ has a very
sharp probability distribution with a large amplitude of the lowest
spin projection, i.e., $m=0$ (for even $N$) or $m=\pm1/2$ (for odd
$N$) \cite{LW10041}. Such a sharp probability distribution of the
SSS can be explained qualitatively by considering the familiar phase
model \cite{phase model} (see also references therein).

\begin{figure}[tbph]
\begin{center}
\includegraphics[width=8.5cm]{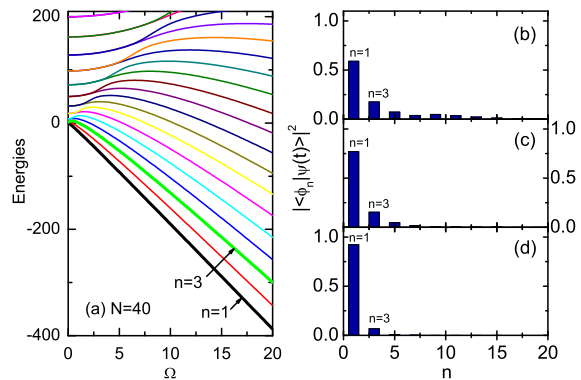}
\end{center}
\caption{(color online) (a) Part of the Eigenenergies
$\mathcal{E}_{n}$ as a function of $\Omega$ for $N=40$. The spectral
distribution $|\langle \protect\phi_n|%
\protect\psi(t)\rangle|^2$ for (b) $\Omega=\protect\kappa$, (c) $\Omega=4.24%
\protect\kappa$, and (d) $\Omega=20\protect\kappa$.} \label{fig5}
\end{figure}

In order to determine the time scale $t_0$, we employ numerical
diagonalization of the Hamiltonian (\ref{hamiltonian}) to obtain a
set of eigenenergies $\{\mathcal{E}_{n}; n=1, 2, 3,...\}$, where
$n=1$ denotes the ground state, $n=2$ the first excited state, and
$n=3$ the second excited state, etc., thus
\begin{equation}
\hat{H}|\phi_n\rangle=
\mathcal{E}_{n}|\phi_n\rangle,\label{eigenEq}
\end{equation}
where $\mathcal{E}_{n}$ and $|\phi_n\rangle$ with $n=1, 2, 3,\cdots,
(2j+1)$ depend on the parameters $N$ and $\Omega$ \cite{Note}. As
shown in Fig.~\ref{fig5}(a), we plot parts of $\mathcal{E}_{n}$ for
$j=20$ ($N=40$) as a function of the coupling $\Omega$. Similar with
previous two- and three-particle cases, we
suppose that the MST $t_0$ depends on the level spacing between $n=1$ and $%
n=3$, namely $t_{0}=\pi/(2S)$ with
$2S=\mathcal{E}_3-\mathcal{E}_1$. To check it, in
Table~\ref{table1}, we compare exactly numerical results of the
time $t_0$ with $\pi/(2S)$ for various $N$ and $\Omega$. Our
diagonalization method gives accurate prediction of the time
$t_0$. We remark that for $N=2$ and $N=3$ cases, both two results
are exactly the same.

\begingroup 
\begin{table*}[tbp]
\caption{Comparison of exactly numerical $t_0$, $\protect\pi/(2S)$, and
analytic results of Eq.~(\protect\ref{t_M}) for different $N$ and $\Omega$.
The times are in units of $(100\protect\kappa)^{-1}$.}
\label{table1}
\begin{center}
\vskip 0.2cm
\begin{tabular}{c|ccc|ccc|ccc}
\hline\hline
&  & $N=40$ &  &  & $N=200$ &  &  & $N=1000$ &  \\ \hline
$\Omega/\kappa$: & 1 & 4.24 & 20 & 1 & 6.7 & 25 & 1 & 10.8 & 50 \\ \hline
Exact num. & 18.28 & 9.065 & 3.604 & 8.192 & 3.184 & 1.549 & 3.665 & 1.104 &
0.4945 \\
$\pi/(2S)$ & 19.02 & 8.615 & 3.573 & 8.143 & 3.048 & 1.533 & 3.571 & 1.071 &
0.4916 \\
Eq.~(\ref{t_M}) & 17.56 & 8.529 & 3.927 & 7.854 & 3.034 & 1.571 & 3.512 &
1.069 & 0.4967 \\ \hline\hline
\end{tabular}%
\end{center}
\end{table*}
\endgroup 

To explain the above agreements, we calculate the spectral
distribution of the spin state, i.e., $|\langle
\phi_n|\psi(t)\rangle|^2$ in Fig.~\ref{fig5}(b)-(d) for $N=40$ and
various $\Omega$. Physically, the spectral distribution measures the
population distribution of the state vector $|\psi(t)\rangle$ on the
eigenstates $|\phi_n\rangle$ \cite{MZI}. For fixed parameters $N$
and $\Omega$, the spectral distribution $|\langle
\phi_n|\psi(t)\rangle|^2$ is time-independent. In fact, one can
expand the spin state in terms of $\{|\phi_n\rangle\}$:
$|\psi(t)\rangle=\sum_n
d_n(t)|\phi_n\rangle$ with the amplitudes $d_n(t)=\exp[-i\mathcal{E}_{n}t]%
d_n(0)$. Here the initial amplitudes $d_n(0)$ depend only on the
initial condition Eq.~(\ref{the initial condition}), therefore the
spectral distribution $|\langle
\phi_n|\psi(t)\rangle|^2=|d_n(t)|^2\equiv|d_n(0)|^2$ and is
time-independent for fixed $N$ and $\Omega$. From our numerical
calculations, Fig.~\ref{fig5}(b-d), we find that total occupation of
the spin state $|\psi(t)\rangle$ on the eigenstates $n=1$ and $n=3$
is over $80$ percent. This is the reason why the MST depends on the
level spacing between these two levels. Moreover, we find the even
$n$ eigenstates are in fact the idle levels, just as previous $N=2$
and $N=3$ cases.

Except for $N=2$ and $N=3$, exact solutions for the nonlinear spin
system within the small-coupling regime ($1<\Omega/\kappa<<N$) do
not exist \cite{Law,Agarwal}. In our previous work \cite{LW10041},
however, we have obtained the analytic expression of the MST based
upon the phase model:
\begin{equation}
\kappa t_0 \simeq \frac{\pi}{2}\sqrt{\frac{\kappa}{2\Omega N}},  \label{t_M}
\end{equation}
which is valid for large $N$ ($\geq10^3$). Our analytic solution
of the MST is derived by the prediction $t_0\simeq T/4$, where
$T=2\pi/\omega_{\text{eff}}$ is the period of the pendulum near
the bottom of a periodic potential \cite{LW10041}. In fact, for
large $N$ the spin system behaviors as a pendulum rotating with
the oscillating frequency
$\omega_{\text{eff}}=\sqrt{2\kappa\Omega_R N}$. As shown in
Table~\ref{table1}, we compare $\pi/(2S)$, the analytic solutions
of Eq.~(\ref{t_M}), and the exact numerical results of the MST for
various parameters $\Omega$ and $N$. It is shown that our
analytical expression of Eq.~(\ref{t_M}) works very well for the
large $N$ ($\sim10^3$), which implies that the oscillating
frequency $\omega_{\text{eff}}$ has its physical meaning to be
half of the level spacing
$S=(\mathcal{E}_3-\mathcal{E}_1)/(2\hbar)$. Note that the phase
model or Eq.~(\ref{t_M}) is valid for the large $N$, while
$\pi/(2S)$ is no limited by this. From this sense, we believe that
the diagonalization method presented here provides much
comprehensive way to measure the maximal-squeezing time.

\section{CONCLUSIONS}

In summary, we have studied the maximal-squeezing time of a
nonlinear spin system, which can be realized in the
two-component BEC, or other spin system similar with Takeuchi et al.\cite%
{Takeuchi}. Motivated by two exactly solvable cases for $N=2$ and
$N=3$, we show that time scale of the maximal squeezing depends on
the level spacing between $n=1$ and $n=3$ eigenstates. We explain it
by calculating the probability distribution of the spin state on the
eigenstates of the Hamiltonian, and find that the above two states
are occupied predominantly. Such results keep valid for arbitrary
$N$ and a wide rage of the coupling strength.

\begin{acknowledgments}
We thank Profs. C.~K. Kim, K. Nahm, C.~P. Sun, W.~M. Liu, S. Yi,
and X. Wang for helpful discussions. This work was supported by
Korea Research Foundation Grant (KRF-2006-005-J02804 and
KRF-2006-312-C00543).
\end{acknowledgments}



\begin{thebibliography}{*}
\bibitem{Note} Except Table I, we choose a fixed
scatting strength $\kappa=1$ throughout our paper, so the time scale
is in units of $\kappa^{-1}$.
\end{thebibliography}
\end{document}